\documentclass[onecolumn]{openjournal}

\usepackage[utf8]{inputenc}   
\usepackage{graphicx}
\usepackage{amsmath,amssymb}
\usepackage{natbib}
\usepackage{subfigure}

\begin{document}
	
	\title{The effect of non-uniformity density structure on the molecular cloud-cores magnetic braking in the Ideal MHD framework}
\author{Abbas Ebrahimi\textsuperscript{*}, Mohsen Nejad-Asghar, Azar Khosravi}

\affil{Department of Physics, University of Mazandaran, Babolsar, Iran\\
	\textsuperscript{*}Corresponding author: \texttt{abbas\_ebri@stu.umz.ac.ir}
}   
   
	\begin{abstract}
		The phenomenon of magnetic braking is one of the significant
		physical effects of the magnetic field in rotating molecular clouds. The physical characteristics of the core can affect on the core rotation rate. and one of the important parameter is the core density structure. According to observation, by regarding the power-law density distribution, $r^{-p}$, for molecular cloud cores, using smoothed particle hydrodynamics simulation, the results show that the increasing of density steepness (i.e., larger $p$) leads to the intensity of the toroidal components of the magnetic field and as a result larger $B_{\phi}$-components lead to more transfer of angular momentum to the outward. Thus, results show that the magnetic braking being stronger with increasing density slope in non-uniform molecular core. For example, the rotation of the system can approximately decrease by fifty percent from $p=0.2$ to $p=1.8$ for a non-uniform system.
	\end{abstract}
	
	.

	\section{Introduction}
	The formation of stars is perhaps one of the least understood and messy processes in cosmic evolution. This process happens in environments that are structured on many scales and are generally referred to as a cloud. Stars are formed in dense regions within the molecular cloud, known as pre-stellar cores, with a density range from $\ge10^5 - 10^6 \mathrm{cm^{-3}}$ at the centre to $\le10^2 \mathrm{cm^{-3}}$  at the edge, embedded in the clumps (e.g.~\citealt{mye83,lad93,wil99,ber07,and09,fer22}).Dense cores observationally have been thoroughly examined from the molecular spectral line emission (e.g.~\citealt{ben89,jij99}), infrared absorption (e.g. \citealt{tei05,lad07,mac17}), and submillimeter dust emission (e.g. \citealt{kirk05,mar16}). The process of gas collapse that leads to the formation of protostars has been studied for decades. Among the early studies in this field, the works of \citet{lar69}, \citet{pen69} and \citet{shu77} can be mentioned. These are generally known as "outside-in" and "inside-out" models, representing the succession of spherical collapse.
	
	Throughout the course of star formation process, various physical mechanisms, such as magnetic field, turbulent, rotation, etc., are influential in the process of star formation and its by-products (i.e., planets). The magnetic field of interstellar environments strongly influences the structure and evolution of the molecular clouds(e.g.~\citealt{hen19}). Observations show that the molecular cloud cores have internal rotations(e.g.~\citet{god93,ho94}), considering the magnetic field incorporated with rotation will play a key role in the collapse of the cloud cores, which is known as the "magnetic braking", that arises from the stress caused by the bending of magnetic field lines (e.g.~\citealt{mes65,bod11}). This mechanism was first investigated by \citet{gil74} to study the transport of angular momentum, from a massive cold gas cloud through
	the hot galactic background, by the propagation of Alfv\'{e}n waves. The magnetic braking effect on the evolution of molecular cloud discs and cores was investigated by some pioneer authors (e.g.~\citealt{mou80,kon87,bas94}). The magnetic braking process is also efficient in the collapse and	fragmentation of molecular clouds, which was first investigated by ~\citet{all03} for a rotating toroidal magnetized core, and
	by \citet{bos07} and \citet{bos09} for the collapse and fragmentation of the
	prolate and oblate molecular clouds. Studies showed that by assuming
	an ideal MHD and with the magnetic field aligned with the rotational
	axis of the core, the magnetic braking disrupts the formation of
	disks around the protostars (e.g.~\citealt{gal06,mel08}). The disk formation depends on two main parameters which control the efficiency of magnetic braking: the ratio of azimuthal and vertical components of the magnetic field at the disk surface, and the Alfv\'{e}n speed in the surrounding medium. The high efficiency of the magnetic braking and disruption of disk formation around the protostars is known as the magnetic braking catastrophe \citet{gal09}.
	Three suggestions are presented to resolve the magnetic braking	catastrophe: (1) non-ideal MHD effects including ambipolar diffusion
	(e.g.~\citealt{mel09,dap12,wur16,zha16,zha18,lam19,wur21}), Ohmic dissipation (e.g.~\citealt{dap10,dap12,tom15,wur16,vay18}), and the Hall effect (e.g.~\citealt{li13,tsu5,mar18,zha21}), (2) misalignment between rotational axis and magnetic field (e.g.~\citealt{joo12,li13,kru13,cod14,lee17,yen21}), and (3) turbulence and the dynamical nature of the environment
	(e.g.~\citealt{mac11,sei12,san12,sei13,wur19}). Therefore, we can say that the effect of magnetic breaking on the formation of protostars and their surrounding disks is completely dependent on the physical characteristics of the core and its surrounding environment, such as density, temperature, turbulence, degree of ionization, geometrical structure of the magnetic field, etc.
	
	One of the important parameter on the magnetic braking catastrophe is the density structure (e.g. \citealt{gir11}). Density structures with radial distribution as $\rho \propto r^{-p}$, $(p > 0)$, are common in systems that have reached intermediate asymptotic states (e.g. \citealt{bar96,liz22}). The observations made of the galactic molecular clouds MonR2 and NGC~6334 show the steeper density profiles in the galactic arms, and this slope increases in the outer regions of the core clouds (\citealt{pir09,sch13}). The observations of low-mass cores typically show density distributions with $p$ varying between 1.5 and 2, resembling a finite-size Bonner–Ebert sphere (e.g. \citealt{mot01,alv01,beu23}). ~\citet{lin22} by examining tracers such as $\mathrm{CH_{3}CCH}$, $\mathrm{CH_{3}OH}$ and etc. at scales of $\leq 0.1~\mathrm{pc}$ obtained a radial profile for the density structure with $p=0.25$ to $p=1.7$. \citet{hun10} also observationally obtained a $p$-value between 1.0 to 1.4 for four cores (Serp-Bolo5, Serp-Bolo1, L158, and Thumbprint Nebula). \citet{kri10} simulated a molecular cloud and showed that the density distribution extended as a power-law tail in the star-forming regions. They predicted values from $p=7/4$ to $p=3/2$ that agree with simulations and observations of star-forming molecular clouds. Observational investigation from the samples of dense cores often suggests slopes $p\le2$ in density profiles for low-mass and high-mass cores (e.g. \citealt{hog00,shi02,you03,beu03,but12,li19}). \citet{tat21} investigated the results of mapping observations containing 107 SCUBA-2 cores in the emission lines of molecules $\mathrm{N_{2}H^+}$,$\mathrm{HC_{3}N}$, and $\mathrm{CCS}$ at $82-94\mathrm{GH}$, according to the given mean HWHM radius and beam radius, suggested a power-law index $p<1.4$ for the radial density profile. As for the power-law behavior of the density in the core of molecular clouds, one situation that lead to gravitational collapse is $\rho \propto r^{-2}$ (e.g. \citealt{shu77, dan19}) , while a state like $\rho \propto r^{-1.5}$ will bring an accretion flow around dense core (e.g. \citealt{hoy41,bon52}). By taking into account the turbulent pressures in the observed linewidth-size scaling, we reach flatter distributions proportional to $\rho \propto r^{-1}$ in non-isothermal gas (e.g. \citealt{mcl97}).
	
	Despite the analytical models presented by assuming a variable density, albeit homologously (e.g. \citealt{gil74,mes79}) and also for example \citet{bas94} who already studied the evolution of a collapsing core with "compact, uniform-density central region, shrinking both in size and mass, surrounded by a 'tail' of matter left behind, in which a near power-law density profile is established", here, we according to the above studies and considering that astrophysical magnetic fields are commonly described using ideal magnetohydrodynamics (MHD), which assumes that the gas is fully ionized, with electrons being tightly coupled to the magnetic field and exhibiting no resistivity, therefore, a tempting question is how does the power law density with different powers affect the magnetic Braking process and core collapse (or, in other words, the formation of a protostar)? For this purpose, in this paper, we consider a power-law density structure with assumption that the magnetic field is proportional to density. We use the PHANTOM simulation code \citep{pri18}, to study the effect of the density structure on the magnetic braking. In Section $\S$~2, we briefly discuss our numerical methods, model assumptions, and initial conditions. Section $\S$~3 contains the results of our calculations, and the discussion with conclusions are presented in Section $\S$~4.
    \section{Method}
    The ideal MHD equations for the fluid of a prestellar core solved by using a smoothed-particle  (magneto)hydrodynamics (SPH) code given by
    \begin{equation}\label{masscounti}
    	\frac{d\rho}{dt}=-\rho \nabla\cdot\textbf{v},
    \end{equation}
    \begin{equation}\label{momentum}
    	\frac{d\textbf{v}}{dt}=-\frac{1}{\rho}\nabla\cdot\left[\left(P+\frac{B^2}{2\mu_0}\right)\mathbb{I}-\frac{\textbf{BB}}{\mu_0}\right],
    \end{equation}
    \begin{equation}
    	\frac{du}{dt}=-\frac{P}{\rho}\nabla\cdot\textbf{v},
    \end{equation}
    \begin{equation}
    	\frac{d\textbf{B}}{dt}=(\textbf{B}\cdot\nabla)\textbf{v}-\textbf{B}(\nabla\cdot\textbf{v}),
    \end{equation}
    where $\frac{d}{dt}\equiv\frac{\partial}{\partial t}+\textbf{v}\cdot\nabla$ is the Lagrangian derivative, $\rho$ is the density, \textbf{v} is the velocity, \textit{u} is the specific internal energy, $\mu_0$ is  the permeability of free space, $\mathbb{I}$ is the identity matrix, P and \textbf{B} are thermal pressure and magnetic field , respectively. The density and smoothing length of each SPH particle are calculated iteratively by summing over its nearest neighbours and using $h = 1.2(\frac{m}{\rho})^{1/3}$ where $h$, $m$ and $\rho$ are the SPH particle’s smoothing length, mass and density, respectively (e.g. \citealt{pri04}). The MHD equations are closed by an equation of state relating the pressure to the density and/or internal energy:
    
    
    \begin{equation}
    	P=(\gamma-1)\textit{u}, 
    \end{equation}
    where $\gamma$ is the adiabatic index.
    
    We set up a dense, cold, spherical, non-uniform density, slowly rotating molecular cloud core of mass $M=1 M_\odot$ and radius $r_0 = 4 \times 10^{16}~\mathrm{cm}$. In this work, we consider, based on observations, the spherical core density, beyond the flattened central region which has a density $\rho_0$\footnote{The central density is calculated according to the Plummer-like model for the inner regions of the core (e.g. \citealp{aud19}).} (corresponding to $n_0=\rho_0/\mu m_H\sim10^{10}\mathrm{cm^{-3}}$), as following:
    \begin{equation}
    	\rho(r)=\rho_0 \left(\frac{r}{r_0}\right)^{-p},
    \end{equation}
    where $r\le r_0$ and $p>0$. This core is placed in pressure equilibrium inside a larger, cubic domain with a side length of $\textit{l} = 8 \times 10^{16} \mathrm{cm}$. The density ratio between the core and the warm medium is assumed to be $30: 1$. The sound speed, by assumption a given temperature $\sim 14 \mathrm{K}$, in the cores is set to $c_s = 2.18 \times 10^4 \mathrm{km \,s^{-1}}$, because the given density profile $\rho\propto r^{-p}$ cause the density to decrease with radius, while this affects the gas pressure, then, the ratio $P/\rho$ remains constant, therefore, the sound speed does not vary accros the core. For simplicity, we use periodic but non-self-gravitating boundary conditions on the global domain (\citealp{wur16}). The core initial rotation is about the z-axis and equal to $\Omega=1.77\times10^{-13} \mathrm{rad \,s^{-1}}$. We assume a magnetic field proportional the density(\citealt{cru12}) as follows:
    \begin{equation}
    	B(\rho)=B_0 \left(\frac{\rho(r)}{\rho(r_0)}\right)^{\beta},
    \end{equation}
    where $B_0$ is a initial magnetic field, $\beta$ is equal to $0.6$, which is the expected scale for conserving magnetic flux during a spherical mass contraction (e.g. see \citealp{shu92,mye21}).
    
    To calculate the angular velocity of the core we considered a shell of the core. By giving velocity ($\textit{v}_x$, $\textit{v}_y$) to all the particles and averaging overall, we obtained an average value for the angular velocity of the shell. We run the models with $4\times10^5$ particles\footnote{By investigating a higher resolutions, finding that it has a negligible impact on our results, about $2\%$}.	
	\section{ RESULTS}
\begin{figure}
		\centering
		\subfigure[]{%
			\includegraphics[width=0.8\textwidth]{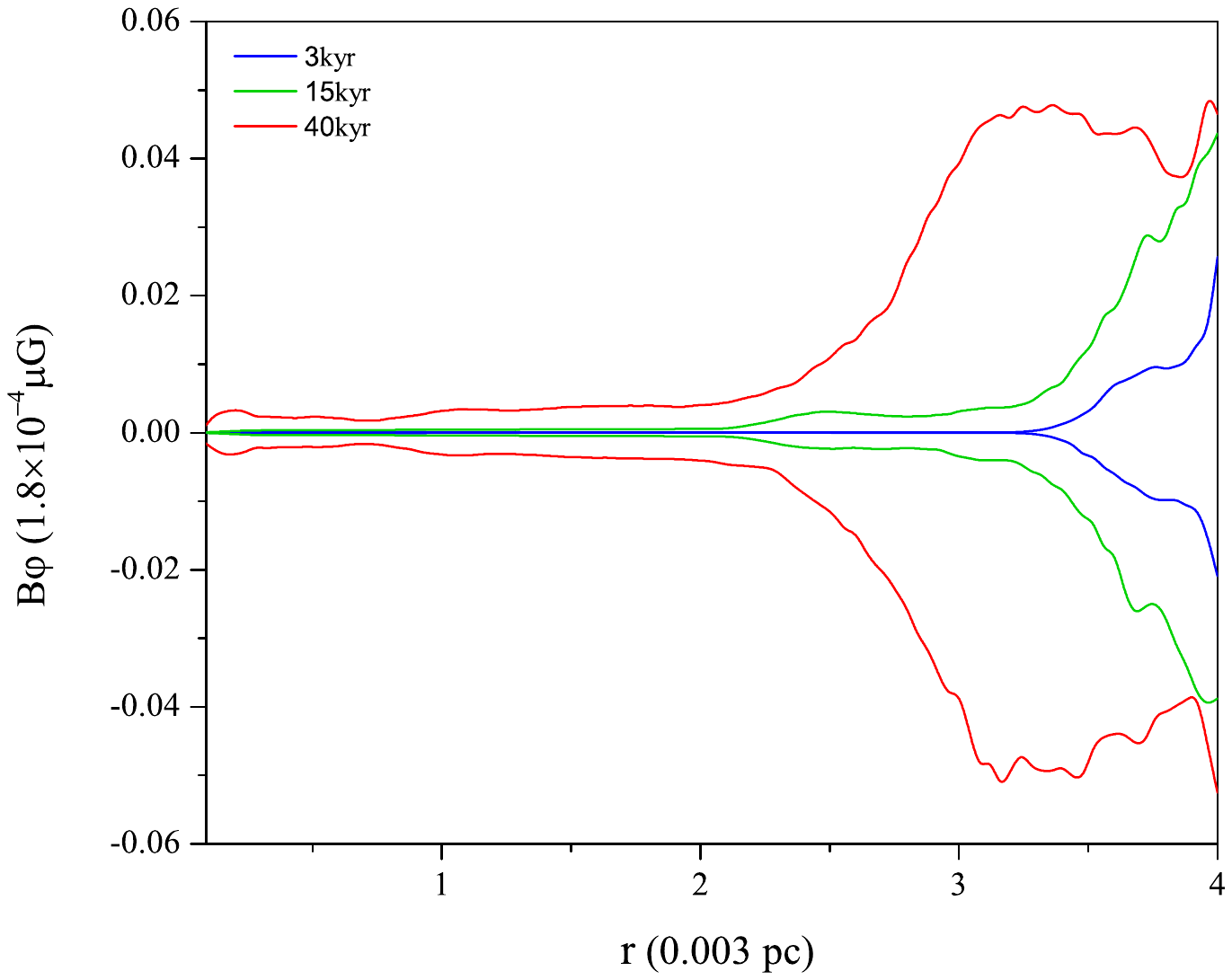}%
			\label{fig:BphiU}%
		}
		\hfill
		\subfigure[]{%
			\includegraphics[width=0.8\textwidth]{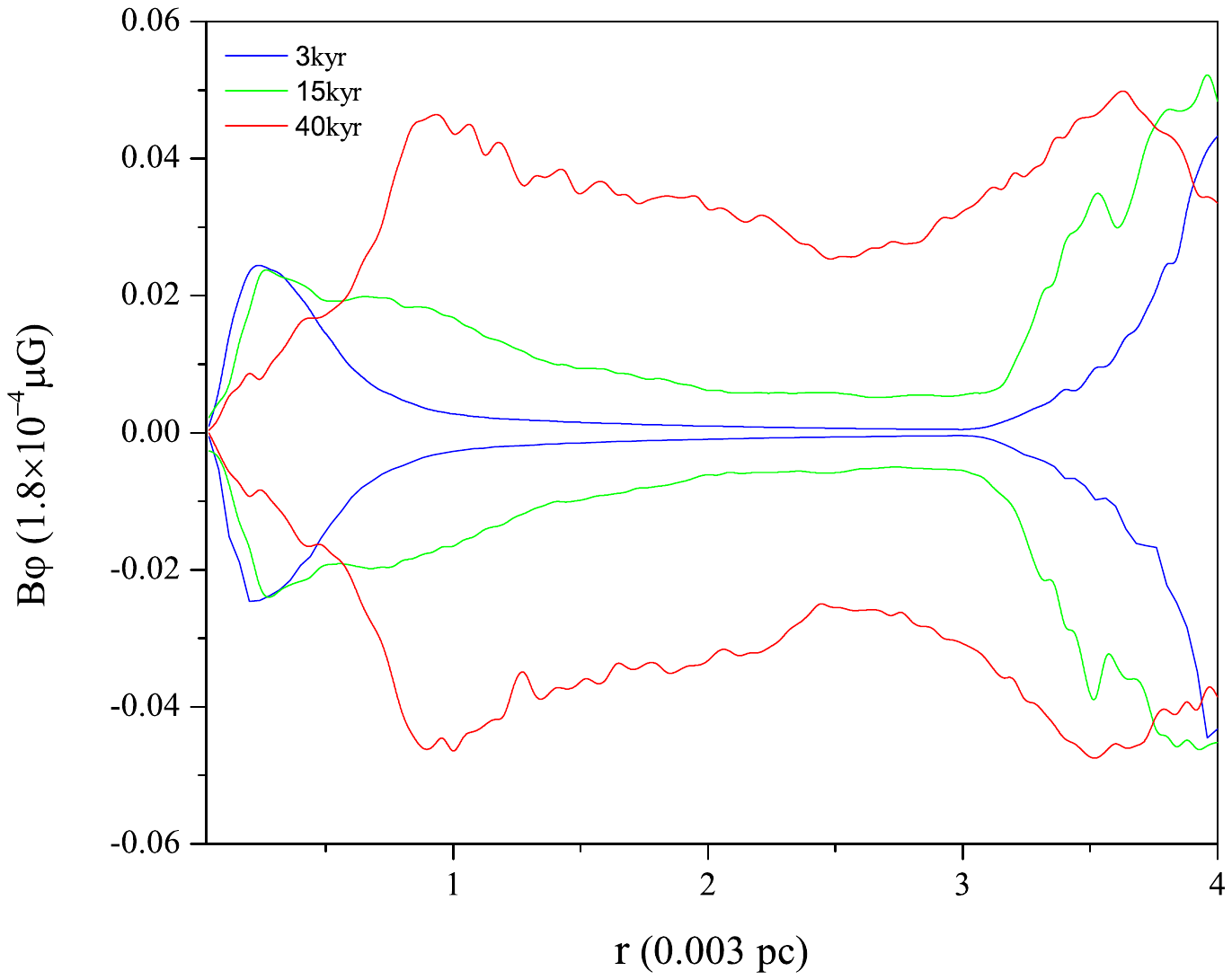}%
			\label{fig:BphiP}%
		}
		\caption{The toroidal magnetic field as a function of radius for initial magnetic field \(B_0=30\,\mathrm{\mu G}\), mass 1\(\textup{M}_\odot\), and three times: \(3\,\mathrm{K\,yr}\) (green), \(15\,\mathrm{K\,yr}\) (blue) and \(40\,\mathrm{K\,yr}\) (red) for (a) a core with uniform density \(\rho_0\) and (b) a core with non-uniform density in power-law model with \(p=1.2\).}
		\label{Bphi-r}
	\end{figure}

	The rotation of the cloud core through the ambient medium bends the magnetic fields, making a $B_{\phi}$, which can introduce a force torque to the material in the opposite direction of the core's rotation. As a result, larger $B_{\phi}$ components lead to more transfer of angular momentum to the outward. In other words, it increases the probability of mass collapse and structure formation. Therefore, in Fig.~\ref{Bphi-r}, the toroidal magnetic field ($B_{\phi}$) is  plotted at different core radii and for three different times $3\mathrm{K\,yr}$, $15\mathrm{K\,yr}$, and $40\mathrm{K\,yr}$ (the period of core rotation is $80\mathrm{k\,yr}$), show the uniform and power-law profile density states, respectively.
	
	As seen in Fig.~\ref{Bphi-r}, the intensity of the toroidal magnetic field in the inner and middle regions of the core with non-uniform density increases significantly compared to the cores with uniform density. Also, a comparison of two Figures shows that the effect of density changes in the production of toroidal magnetic field in the middle regions of the core appears more in longer times. For example, in $3\mathrm{k\,yr}$, the magnetic field in the middle regions is not much different from a core with a uniform density, but after half the core rotation time, the difference between the magnetic fields in the uniform and non-uniform state is very prominent.
	
	\begin{figure}[ht!]
		\centering\includegraphics[scale=0.5]{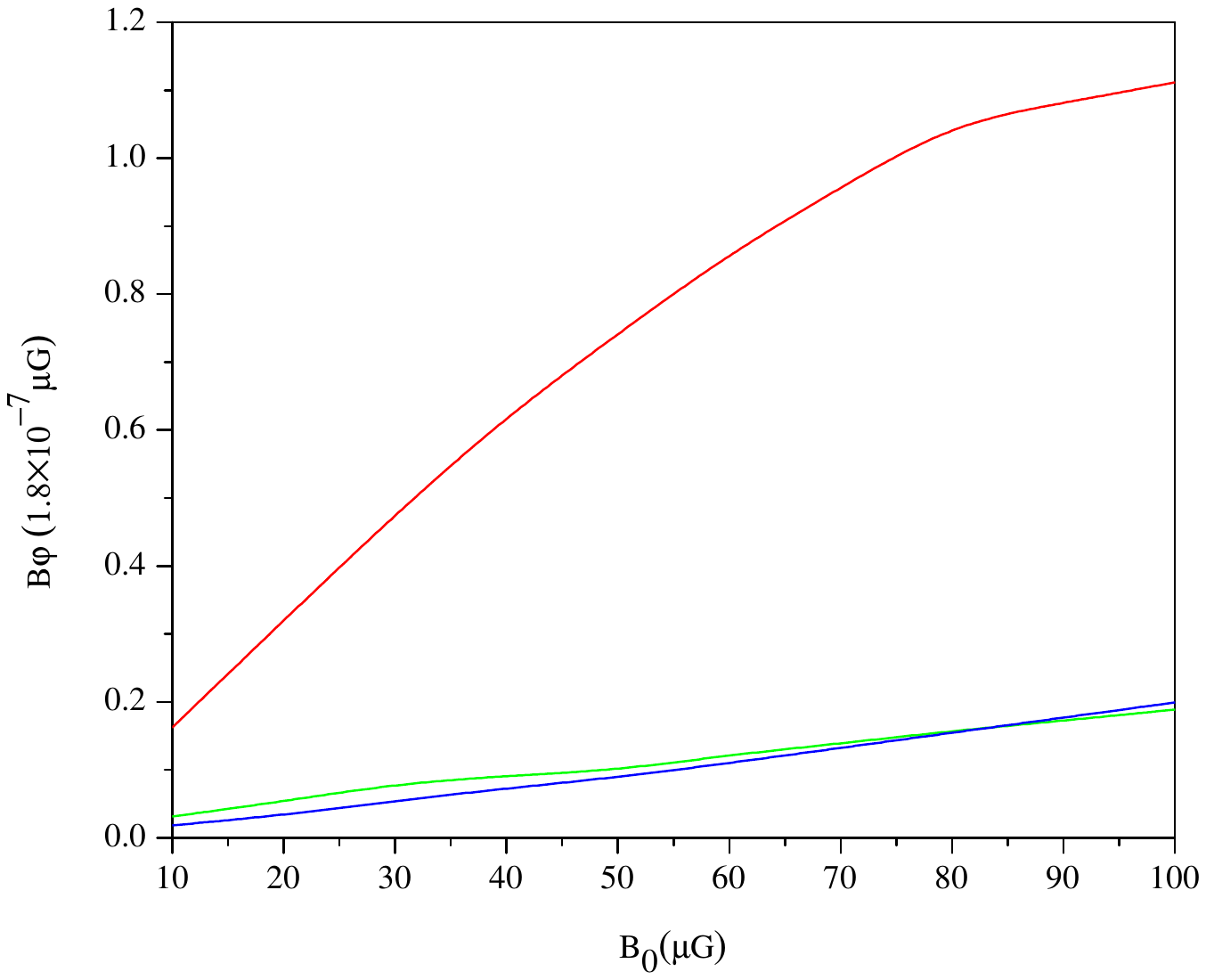}
		\caption{The toroidal magnetic field, $B_{\phi}$, as a function of different initial magnetic field values, $B_0$, for a 1\(\textup{M}_\odot\) core with power-law index $p=1.2$ after $15\mathrm{K\,yr}$ time, at radii of $5.4\times 10^{-3}$ (blue), $8.4\times 10^{-3}$ (green), and $11.4\times 10^{-3}  \mathrm{pc}$ (red) .}\label{BzeroVsBphi}
	\end{figure}
	
	\begin{figure}[ht!]
		\centering\includegraphics[scale=0.5]{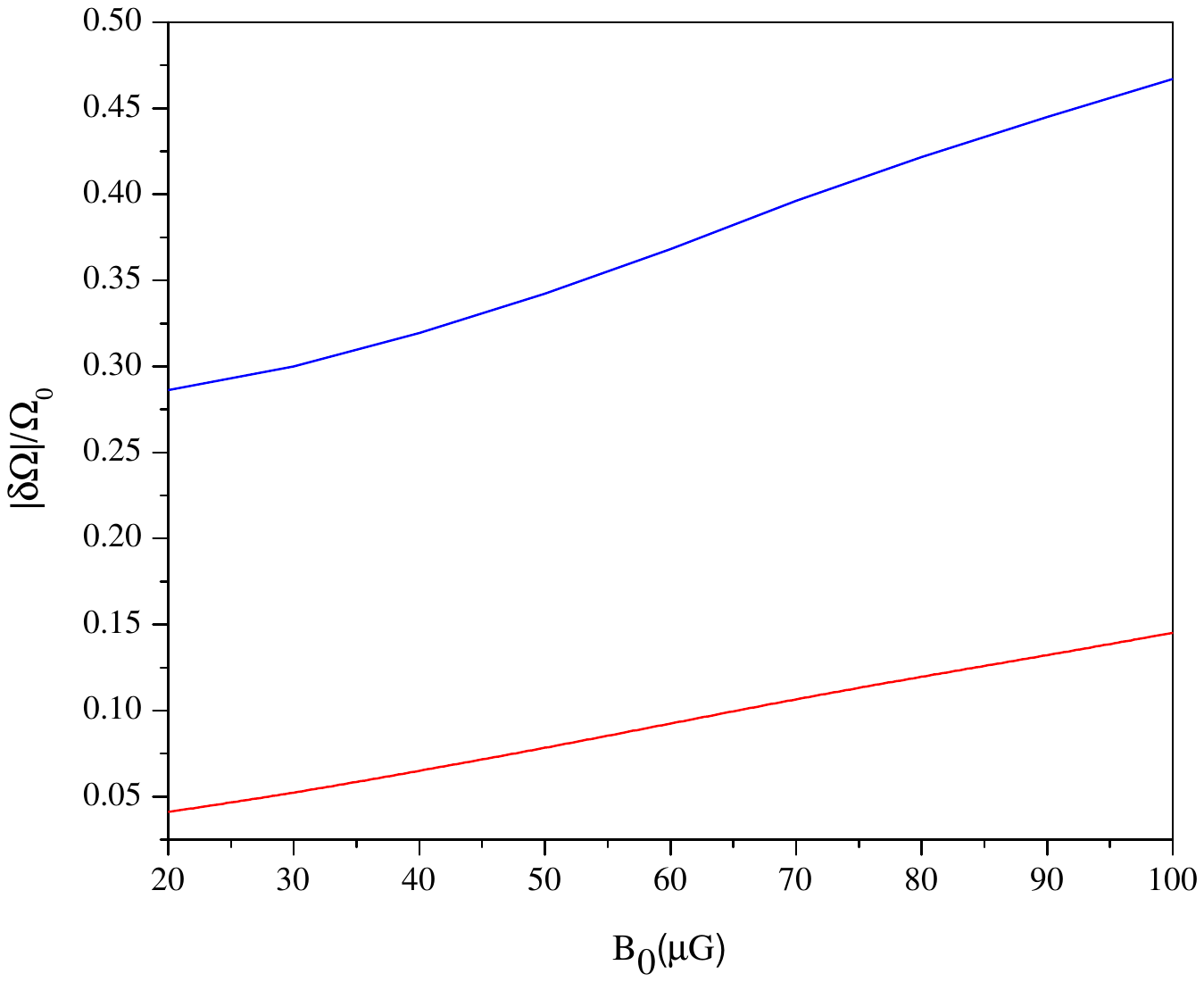}
		\caption{The relative changes of the rotation for a uniform (red) and non-uniform (blue) density, with $p$-index $1.2$, core as a function of initial magnetic field, $B_0$, over $40\mathrm{K\,yr}$ .}\label{co}
	\end{figure}
	
	\begin{figure}[ht!]
		\centering\includegraphics[scale=0.5]{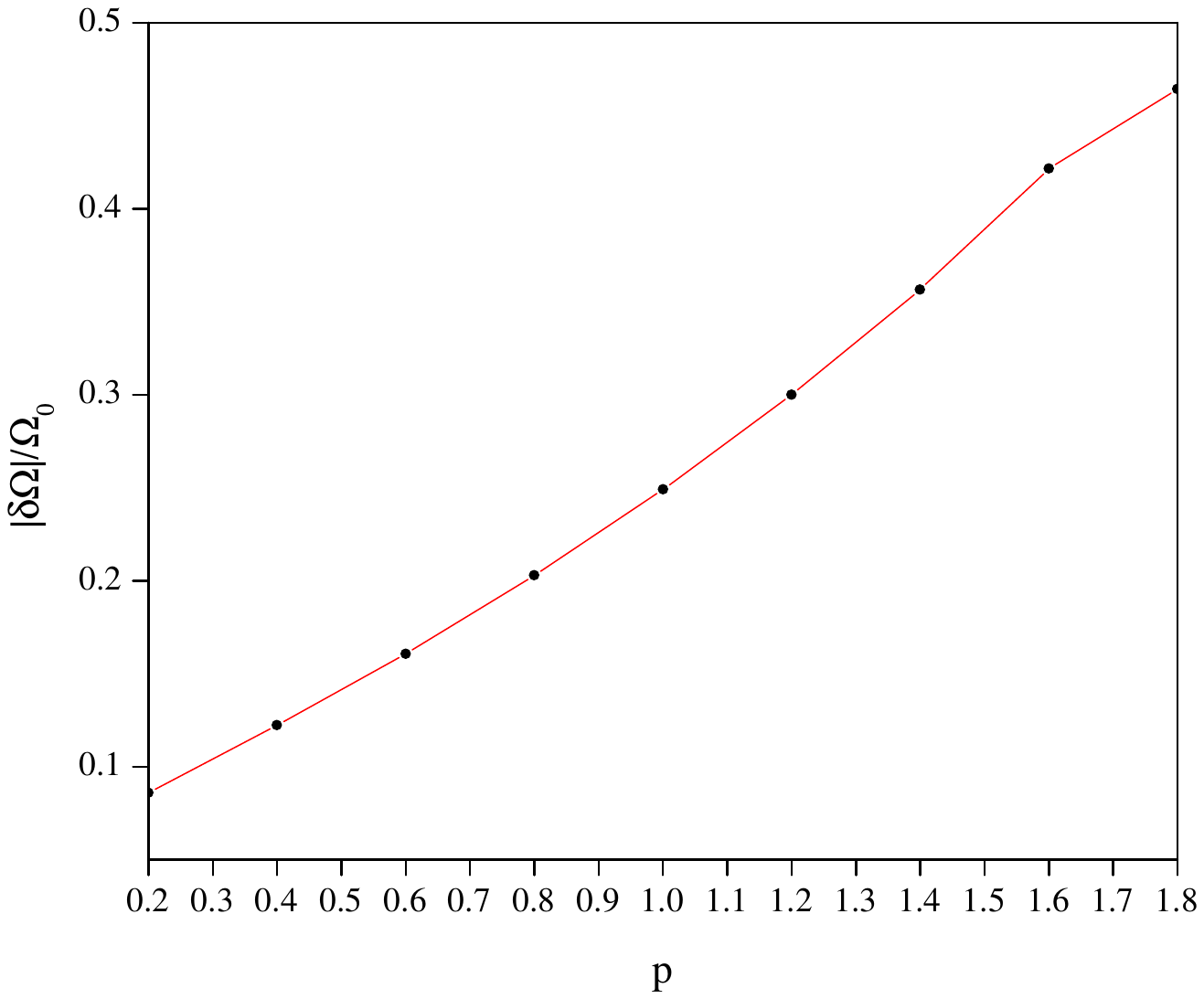}
		\caption{The relative changes of the rotation for a non-uniform density core, with $B_0=30\mathrm{\mu\,G}$ and mass 1\(\textup{M}_\odot\), as a function of different values of $p$-index at $40\mathrm{K\,yr}$.}\label{op}
	\end{figure}
	
	In Fig.\ref{BzeroVsBphi} we plot the toroidal magnetic field versus $B_0$ (initial magnetic field). It is seen that the increasing $B_0$ leads to increase of toroidal magnetic field component. We know that the toroidal magnetic field and its intensity have a direct effect on the magnetic braking and rotation rate of core. 
	In Fig.\ref{co} we show that the rotation rate of core for different values of initial magnetic field for uniform and non- uniform density structures. It is seen that increasing the magnetic field leads to a decrease in the rotation. Also, as we can see in the Fig.\ref{co}, in the system with power law density model, the reduction of rotation is more than the system with uniform density model.
	
\begin{figure}[ht!]
		\centering\includegraphics[scale=0.5]{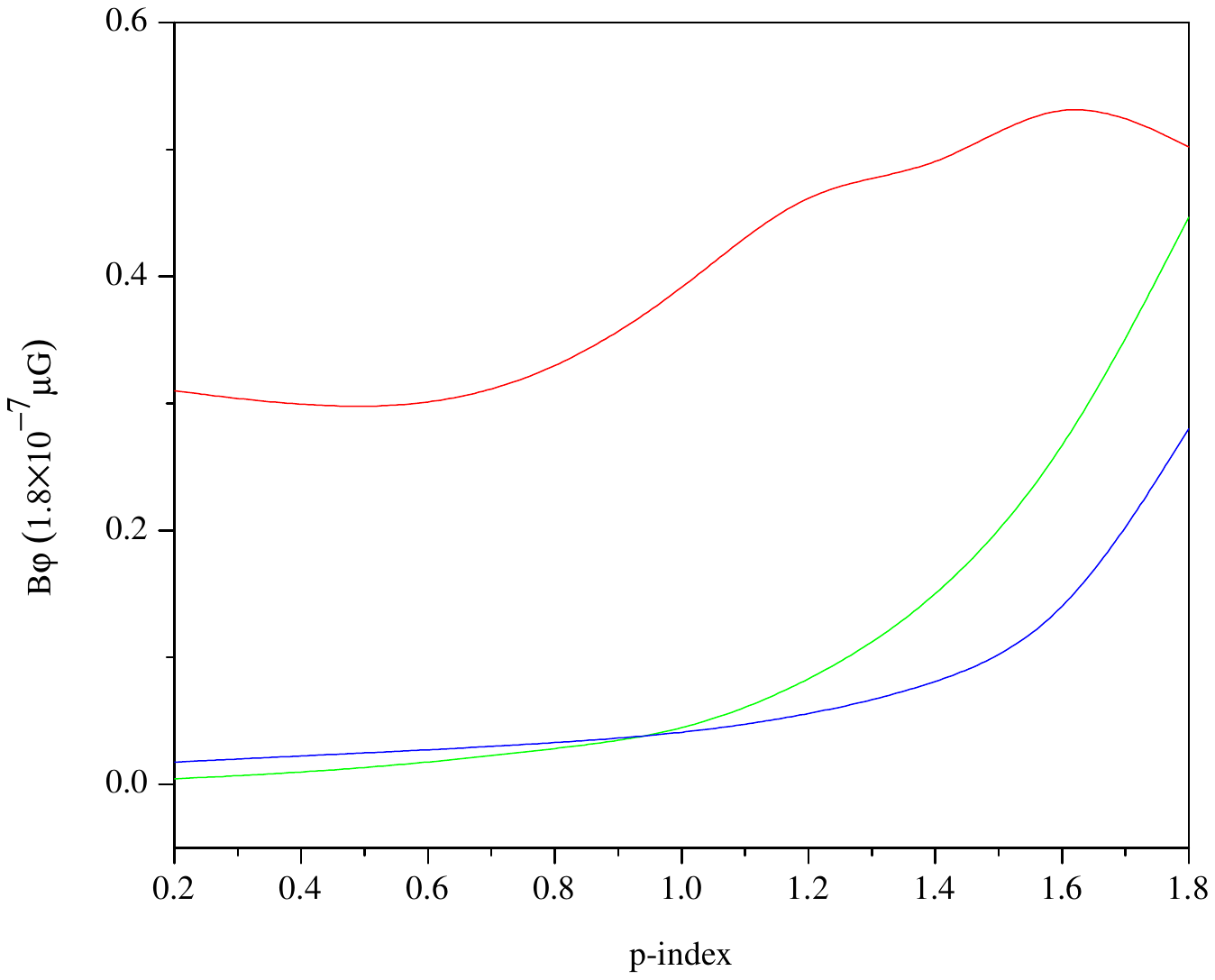}
		\caption{The toroidal magnetic field, $B_{\phi}$, as a function of different $p$-index values, for a 1\(\textup{M}_\odot\) core  over a $15\mathrm{K\,yr}$ period, shown for radii of $5.4\times 10^{-3}$ (green) $8.4\times 10^{-3}$ (green), and $11.4\times 10^{-3}  \mathrm{pc}$ (red).}\label{BphiVsP}
	\end{figure}

	Noted that, the power index for power-law density typically is taken as $p=1.2$. The results show that, on average, the difference between two mentioned models is approximately $25$ percent. At follow, we examined the power law density model with different values of $p$-index. Results are shown in Fig.\ref{op}. It is seen that power law models with larger p values lead to faster reduction of core rotation. It can be concluded that the magnetic braking effect increases in such models. The toroidal magnetic field has a direct effect on the magnetic braking, so we can conclude that the increasing $p$-index leads to an increase in the toroidal magnetic field. This result is also shown in Fig.\ref{BphiVsP}.
	
	\section{Conclusion}
	In this paper, we studied the magnetic braking effect in the core of quasi-static molecular clouds. To investigate this issue, we used SPH numerical simulation. In the studied models, it was assumed that the density of the cores is uniform. However, there is observational evidences that the density inside the core can be non-uniform (e.g. \citealp{bar96}). This led us to investigate the effect of the non-uniform density of the molecular cloud core on the magnetic braking effect. For this purpose, we considered a power-law model for the core density as $\rho(r)=\rho_0 (r/r_0)^{-p}$ where the $p$-index in a way indicates the degree of non-uniformity. 
	
	According to the results, we found that the toroidal magnetic field in non-uniform systems has larger values than in the uniform density case, and increasing p (the amount of non-uniformity) further increases the toroidal magnetic field intensity (Fig.\ref{Bphi-r} and \ref{BphiVsP}). Also, the result show that the toroidal magnetic field have larger value in the outer radius of core in uniform density system. However, it can be said that in systems with non-uniform density, the magnetic field has a significant value in all radii. The Fig.\ref{BzeroVsBphi} Show that the value of toroidal magnetic field is dependent on the initial magnetic field of cloud, $B_0$. The bigger value of $B_0$, leads to bigger toroidal magnetic field in the core. According to \citealp{cru12} we know that $B \propto \rho^\beta$ where $\beta=0.6$. Clouds with greater value of density have a larger initial and so larger toroidal component magnetic field. 
	
	In Fig.\ref{Bphi-r}, for a uniform initial density, we see that the magnetic field in the outer regions increases with time. Therefore, we can conclude that after a certain period of time, the uniformity of the density of the system is lost and the outer regions have higher densities than the inner regions. However, as can be seen for non-uniform model, the magnetic field in the inner and outer regions does not differ much, and the gradient of the field becomes smoother with the radius. Therefore, according to the relationship between field and density, we can conclude that after a certain period of time, the density gradient of the system becomes smoother too. On the other hand, as we have already mentioned, the increase of the toroidal magnetic field leads to the increase of magnetic braking effect. This is because of the magnetic lined become twisted and the twist propagate outward along the field line at the Alfven velocity. The barking time is given as $t_b \propto \rho/V_A$ where $V_A$ is the Alfv\'{e}n speed that is proportional to $B^2$ and the analytical study showed that $\Omega(t)\propto exp(-t/t_b)$ (e.g. \citealp{bod11}). Therefore, the rotation speed of the system should be reduced.
	
	In Fig.\ref{co}, the relative changes of the rotation speed are plotted in terms of the initial field for uniform (red line) and non-uniform density (blue line). It can be seen that the rotation of the system or in other words the angular momentum of the system in the non-uniform state is significantly reduced compared to the uniform state. The results show that, on average, the difference between two mentioned models is approximately \%$25$ for $p=1.2$. Also, Fig.\ref{op} shows that if the degree of non-uniformity of the system, $p$, increases, the rate of decline of the rotation of the system increases. For example, the rotation of the system can approximately decrease by \%$50$ from $p=0.2$ to $p=1.8$ for a non-uniform system.
	
	In general, it can be said that in the uniform density case, the $B_{\phi}$ magnetic field components appear in the outer regions of the core, and the decrease in core rotation in this case is less than in the non-uniform case where we have a $B_{\phi}$ component throughout the core. The rate of decrease in core rotation in the non-uniform model depends on $p$, and in this presented model, for $p=1.2$, about \%$25$ decreased. For the power-law density profile, the potential and gravitational force change with distance much more slowly than for a point mass. This could indicate that the environment of the cores is probably gravitationally bound to them (e.g. \citealp{kirk17}) that in turn could be a confirmation of more efficiency on magnetic braking.

\end{document}